\documentclass[aps,prl,twocolumn,groupedaddress,showpacs]{revtex4} 

\usepackage{graphicx}
\usepackage{dcolumn}
\usepackage{bm}

\begin{document} 

\title{Metal-to-Insulator Crossover in YBa$_2$Cu$_3$O$_y$ 
Probed by Low-Temperature Quasiparticle Heat Transport} 

\author{X. F. Sun} 
\email[]{ko-xfsun@criepi.denken.or.jp} 
\author{Kouji Segawa} 
\author{Yoichi Ando} 
\email[]{ando@criepi.denken.or.jp} 
\affiliation{Central Research 
Institute of Electric Power Industry, Komae, Tokyo 201-8511, 
Japan.} 

\date{\today} 

\begin{abstract} 

It was recently demonstrated that in La$_{2-x}$Sr$_x$CuO$_4$ the 
magnetic-field ($H$) dependence of the low-temperature thermal
conductivity $\kappa$ up to 16 T reflects whether the normal 
state is a metal or an insulator. We measure the $H$ dependence 
of $\kappa$ in YBa$_2$Cu$_3$O$_y$ (YBCO) at subkelvin 
temperatures for a wide doping range, and find that at low doping 
the $\kappa(H)$ behavior signifies the change in the ground state 
in this system as well. Surprisingly, the critical doping is 
found to be located deeply inside the underdoped region, about 
the hole doping of 0.07 hole/Cu; this critical doping is 
apparently related to the stripe correlations as revealed by the 
in-plane resistivity anisotropy.

\end{abstract} 


\maketitle 

It is well recognized that the high-$T_c$ superconductivity in 
cuprates is realized in a strongly-correlated electron system 
that is essentially a doped Mott insulator. Studying the 
low-temperature normal state, the ground state in the absence of 
superconductivity, is important in understanding the rather 
mysterious electronic state from which the superconductivity 
emerges. In the past, the low-temperature normal state has been 
investigated by measuring the resistivity while destroying the 
superconductivity with a very high magnetic field in systems that 
have relatively low $T_c$, such as La$_{2-x}$Sr$_x$CuO$_4$ (LSCO) 
\cite{Ando1,Boebinger}, Bi$_2$Sr$_{2-x}$La$_x$CuO$_{6+\delta}$ 
(BSLCO) \cite{Ono}, and Pr$_{2-x}$Ce$_x$CuO$_4$ (PCCO) 
\cite{Fournier}. In these experiments, it was revealed that the 
metal-to-insulator crossover (MIC) in the low-temperature normal 
state takes place at optimum doping (hole doping per Cu, $p$, of 
$\sim$0.16) in LSCO \cite{Boebinger} and in PCCO \cite{Fournier}, 
while it occurs in the underdoped region (specifically, at $p 
\sim 1/8$) in BSLCO \cite{Ono}. Thus, although the critical doping 
for the MIC was found to be nonuniversal, it was established that 
there is a certain range of doping where the superconductor has an 
``insulator" as its normal state under high magnetic fields. This 
insulator is unusual in that it shows a peculiar $\log (1/T)$ 
divergence in resistivity \cite{Ando1,Boebinger,Ono}.

Since it is exotic to have an ``insulating" normal state in a 
superconductor, and furthermore the properties of the insulator 
are unusual, it is expected that the nature of this insulator 
bears a key to the microscopic understanding of the electronic 
state of the cuprates. Given that the critical doping for the MIC 
is different for LSCO and BSLCO, it is clearly desirable to study 
the MIC in other cuprate systems and gain insights into the cause 
of the insulating behavior. Obvious targets for such studies are 
YBa$_2$Cu$_3$O$_y$ (YBCO) and Bi$_2$Sr$_2$CaCu$_2$O$_{8+\delta}$ 
(Bi2212), but the upper critical fields of these materials are so 
high that it has been difficult to observe the MIC at low 
temperatures. 

Recently, we have demonstrated that the low-temperature 
quasiparticle (QP) heat transport measured in 16-T field can be 
an alternative tool to probe the MIC \cite{Sun}; namely, we found 
in LSCO that whether the low-temperature normal state under 60 T 
\cite{Boebinger} is metallic or insulating corresponds exactly to 
whether the QP heat transport at very low temperature is enhanced 
or suppressed by magnetic fields up to 16 T. (Essentially the 
same result was reported independently by Hawthorn {\it et al.} 
\cite{Hawthorn}.) It was discussed \cite{Sun} that the 
magnetic-field suppression of $\kappa$ at very low temperature is 
related to the magnetic-field-induced spin density wave (SDW) 
\cite{Lake,Khaykovich} or charge density wave (CDW) 
\cite{Hoffman}, that appear to be competing with the 
superconductivity and would localize the QPs, thereby leading to 
the insulating state. It has been proposed \cite{Kivelson} that 
these magnetic-field-induced orders may be related to charge 
stripes \cite{Carlson}, which have been shown to be relevant to 
the charge transport in cuprates at least in the lightly 
hole-doped region \cite{Ando2,Ando3,anisotropy,Ando5}. In fact, a 
recent theory showed \cite{Takigawa} that the 
magnetic-field-induced incommensurate antiferromagnetism can be 
responsible for the localized QP transport at low temperature. In 
any case, the result in LSCO tells us that the magnetic-field 
dependence of $\kappa$ at low-temperatures in YBCO would clarify 
where the MIC lies in this 90-K-class superconductor. The YBCO 
system is particularly useful for examining the relevance of the 
charge-stripe instability, because this is the only system other 
than the La-based cuprates where the neutron scattering has found 
static charge stripes \cite{Mook}.

In this Letter, we locate the MIC in YBCO according to the above 
strategy. We find that the critical doping is unexpectedly low, 
at the oxygen content $y$ of $\sim$6.55, which corresponds to $p 
\sim 0.07$. Interestingly, this low critical doping is exactly 
the same as the boundary below which the charge stripes start to 
dominate the charge transport, as evidenced by our study of the 
in-plane resistivity anisotropy in this material 
\cite{anisotropy}. This result gives evidence that the insulating 
normal state of the cuprate under high magnetic field is related 
to the self-organization of the electrons, perhaps into the form 
of disordered stripes. 

High-quality YBa$_2$Cu$_3$O$_y$ single crystals are grown in 
Y$_2$O$_3$ crucibles by a conventional flux method. The crystals 
are carefully annealed to control the oxygen content and then 
perfectly detwinned \cite{Segawa1}. Here we study untwinned 
single crystals with the oxygen content $y$ of 6.45, 6.50, 6.60, 
6.65, 6.70 and 7.00, for which the zero-resistivity $T_c$'s are 
20, 39, 56, 59, 62 and 91 K, respectively. We emphasize that, to 
reliably control the oxygen content in the lightly doped regime, 
we have established elaborate procedures whose details are 
described in Ref. \cite{Segawa2}. To probe the QP heat transport 
in the CuO$_2$ planes, the thermal conductivity $\kappa$ is 
measured along the $a$ axis, with which the additional electronic 
heat transport coming from the Cu-O chains (that are along the 
$b$ axis) can be avoided. Rectangular-shaped samples with a 
typical size of $1.5 \times 0.5 \times 0.15$ mm$^3$ are used for 
the measurements, where the longest (shortest) dimension is along 
the $a$ ($c$) axis. The measurements in the millikelvin region is 
done by a conventional steady-state ``one heater, two 
thermometer" technique in a dilution refrigerator \cite{Takeya}. 
The magnetic-field dependence of $\kappa$ is measured by the same 
method in a $^3$He refrigerator from 0.36 to 7 K. The magnetic 
field up to 16 T is always applied along the $c$ axis. All the 
$\kappa(H)$ data are taken with the field-cooled procedure 
\cite{Sun,Ando6} to avoid the vortex-pinning-related hysteresis 
\cite{Aubin}. 

\begin{figure} 
\includegraphics[clip,width=8.5cm]{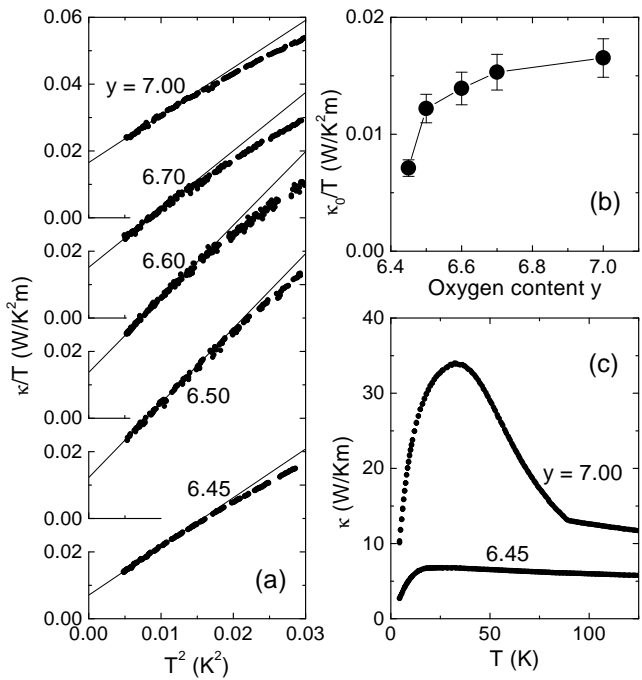} 
\caption{(a) Temperature dependence of the in-plane thermal 
conductivity $\kappa$ in 0 T for YBa$_2$Cu$_3$O$_y$ single crystals 
with various $y$ values. Thin solid lines are fits of the lowest-$T$ 
data to $\kappa/T = a + bT^2$.  
(b) $y$ dependence of the residual QP component, $\kappa_0/T$.  
(c) High-temperature $\kappa(T)$ data for $y$ = 7.00 and 6.45.} 
\end{figure} 

Figure 1(a) shows the thermal conductivity of the YBCO single 
crystals at very low temperatures down to 70 mK. At low enough 
temperature (below $\sim$120 mK), all the data are well described 
by $\kappa/T = a + bT^2$ [see the thin solid lines in Fig. 1(a)], 
where the $T$-linear and cubic dependences of $\kappa$ are 
attributed to the electron (or QP) term $\kappa_e$ and the phonon 
term $\kappa_{ph}$, respectively \cite{Takeya,Taillefer}. Thus, 
the zero-temperature intercept of the straight lines in Fig. 1(a) 
directly gives the value of the residual QP component, 
$\kappa_0/T$. This component $\kappa_0/T$ is known to be due to 
the existence of extended zero-energy QP excitations near the 
nodes of the $d$-wave gap in the presence of even a small amount 
of disorder \cite{Graf,Durst}, and is known to be ``universal" 
against the change in the impurity concentrations 
\cite{Taillefer}. Note that $\kappa_0$ is a good approximation of 
$\kappa_e$ only at very low $T$, since $\kappa_e$ quickly grows 
with $T$ \cite{Graf,Hill}. As shown in Fig. 1(b), $\kappa_0/T$ 
increases gradually with carrier doping, which is similar to the 
behavior in LSCO \cite{Takeya} and is consistent with the 
previous YBCO result \cite{Sutherland}. The smooth evolution of 
the residual QP thermal conductivity is compatible with a robust 
$d$-wave pairing symmetry in the whole doping range; namely, our 
data do not give any hint of a phase transition (as a function of 
$y$) \cite{Dagan} that causes a change in the pairing symmetry 
and removes the nodes from the $d_{x^2-y^2}$-wave gap. 

\begin{figure*} 
\includegraphics[clip,width=17.5cm]{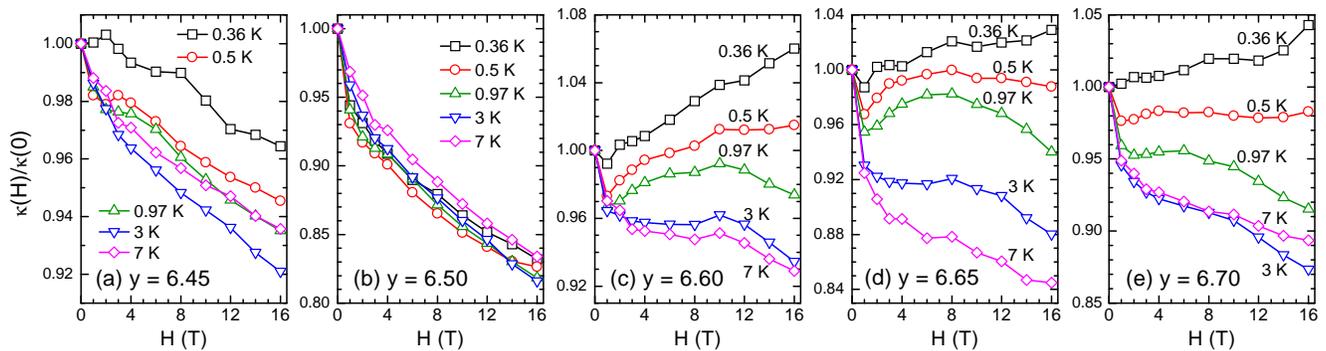} 
\caption{Magnetic-field dependences of $\kappa$ for underdoped 
YBCO single crystals.  Note the distinct behavior at the lowest 
temperature for $y \le 6.50$ and for $y \ge 6.60$.} 
\end{figure*} 

The high-$T$ thermal conductivity for the two limiting 
compositions of the present study, $y$ = 7.00 and 6.45, are shown 
in Fig. 1(c). These data are taken by using a Chromel-Constantan 
thermocouple in a $^4$He cryostat. The peak value (34 W/Km) of 
our $y=7.00$ sample, which reflects the QP lifetime and is a 
measure of the cleanliness of the crystal \cite{Zhang}, is 
considerably larger than the value ($\sim$25 W/Km) for ordinary 
YBCO crystals grown in YSZ crucibles \cite{Hill,Zhang} and is 
close to the value ($\sim$40 W/Km) for the best crystals grown in 
BaZrO$_3$ crucibles \cite{Hill,Zhang}. This result demonstrates 
that our crystals are very clean, which is also evidenced in the 
electric transport data \cite{Segawa1}. The $y=6.45$ sample shows 
little anomaly at $T_c$ (= 20 K), a behavior similar to that of 
underdoped LSCO \cite{Sun}.

The main result of the present work is shown in Fig. 2, where the 
magnetic-field dependences of $\kappa$ is shown for all the 
underdoped samples. For optimally-doped (or slightly overdoped) 
YBCO, it is well known that $\kappa$ is enhanced with $H$ at low 
enough temperatures, but is suppressed with $H$ at higher 
temperatures \cite{Hill,Chiao}; such a behavior is consistent 
with the behavior of optimally-doped Bi2212 \cite{Ando6,Aubin} 
and LSCO \cite{Sun}. Figures 2(c-e) demonstrate that such a 
behavior is also observed in underdoped samples with $y$ = 
6.60--6.70; from these data, one can conclude that there is 
essentially no qualitative change in $\kappa(H)$ down to $y$ = 
6.60. However, upon further reducing the doping level, quite 
strong suppression in $\kappa$ with increasing $H$ suddenly shows 
up in the $y$ = 6.50 sample at 0.36 K [Fig. 2(b)]. The 
suppression of $\kappa$ with $H$ is also observed at $y$ = 6.45. 
Therefore, there is a qualitative change in $\kappa(H)$ at low 
temperature across $y \simeq 6.55$, above which $\kappa$ increases 
with $H$ but below which it decreases with $H$. Note that the 
qualitative change in $\kappa(H)$ is most certainly governed by 
the QP contribution to the heat transport, because it is unlikely 
that the {\it magnetic-field dependence} of the other 
contributions (phonons and magnetic excitations) suddenly change 
sign with doping. Note also that the smooth evolution of 
$\kappa_0/T$ shown in Fig. 1(b) demonstrates that there is no 
sudden change in the pairing symmetry between $y$ = 6.50 and 6.60, 
which means that the distinct change in the low-$T$ $\kappa(H)$ 
behavior is not due to a difference in the superconducting gap 
structure \cite{Ando6}. Thus, as in the case of LSCO 
\cite{Sun,Hawthorn}, the magnetic-field-induced enhancement or 
suppression of the QP heat transport at very low $T$ reflects 
whether the electronic system is a thermal metal or a thermal 
insulator in high magnetic field, which in turn signifies whether 
the high-field normal state is a metal or an insulator. Hence, we 
can conclude that the MIC in YBCO occurs at $y \simeq 6.55$, 
which is deep in the underdoped regime and corresponds to $p 
\simeq 0.07$ according to our systematic study of the Hall 
coefficient \cite{Segawa2}.

One may notice in Fig. 2(a) that the magnetic-field-induced 
suppression of $\kappa$ becomes weaker with lowering temperature, 
which hints at the possibility that $\kappa$ for $y$ = 6.45 might 
eventually increase with $H$ at further lower temperature. To 
confirm that this is not the case, we have measured $\kappa$ for 
$y$ = 6.45 in 16 T down to 100 mK and compared it to that in 0 T. 
As is shown in the inset to Fig. 3(a), $\kappa(16{\rm 
T})/\kappa(0{\rm T})$ is always less than 1 and is declining with 
decreasing $T$, which assures that the magnetic-field-induced 
insulating state for $y \le 6.50$ concluded from Fig. 2 is indeed 
a zero-temperature property. The relative smallness of the $H$ 
dependence in Fig. 2(a) is probably because the QP contribution 
to the total $\kappa$ is the smallest in this most underdoped 
sample [see Fig. 1(b)].

It is useful to note that the data in the inset of Fig. 3(a) help 
us to estimate how high a magnetic field is needed to induce a 
true insulator with zero electrical conductivity: The data 
indicate that $\kappa$ for $y$ = 6.45 is reduced by $\sim$15\% in 
16 T at 100 mK. At this temperature, the fitting for $y$ = 6.45 
in Fig. 1(a) tells us that $\kappa$ is well described by $\kappa 
/T = 7.32\times 10^{-3} + 1.41T^2$, where the first (second) term 
comes from $\kappa_e$ ($\kappa_{ph}$). Therefore, one can 
confidently estimate $\kappa_e / \kappa$ to be $\sim$0.34 at 100 
mK, and this estimate indicates that the 15\% reduction of 
$\kappa$ in 16 T is already halfway towards the complete 
suppression of $\kappa_e$; this suggests that in the magnetic 
field of order 50 T the electronic contribution to $\kappa$ would 
be quenched ({\it i.e.} a true insulator is achieved), which 
seems to be reasonable for a sample with $T_c$ of 20 K 
\cite{Ando1}.

\begin{figure} 
\includegraphics[clip,width=6.5cm]{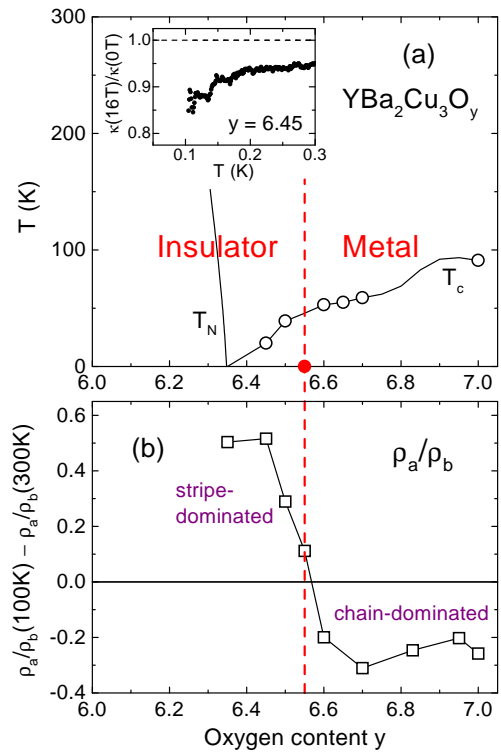} 
\caption{(a) Phase diagram of YBCO.  $T_N$ and $T_c$ indicate 
the  N\'eel temperature \cite{Ando2} and the superconducting 
transition temperature \cite{Segawa1}, respectively. Open circles 
show  $T_c$'s of the crystals studied in this work.  The 
metal-insulator crossover in the normal state, determined from 
the low-$T$ $\kappa(H)$ behavior, occurs at $y \simeq 6.55$ and 
is shown by a dashed line. Inset: $T$-dependence of the ratio 
$\kappa(16{\rm T})/\kappa(0{\rm T})$ for $y$ = 6.45  down to 100 
mK. (b) Difference between $\rho_a/\rho_b$ at 100 K and that at 
300 K; data are taken from Ref. \cite{anisotropy}.} 
\end{figure} 

Based on the above observations, one can draw a phase diagram of 
YBCO as depicted in the main panel of Fig. 3(a), where the MIC 
lies at $y \simeq 6.55$. Interestingly, in the light of the 
unusual doping-dependence of the in-plane resistivity anisotropy 
in YBCO \cite{anisotropy}, this value of the critical doping 
gives us a new insight into the nature of the MIC: As discussed 
in the introduction, one of the likely candidates for the cause 
of the QP localization observed in underdoped LSCO is the 
magnetic-field-induced stabilization of the charge stripes, if the 
static stripe state is competing with superconductivity 
\cite{Kivelson}. In Ref. \cite{anisotropy}, it was demonstrated 
that in nearly optimally-doped YBCO the in-plane resistivity 
anisotropy ratio $\rho_a/\rho_b$ has just a weak temperature 
dependence at high temperature and decreases at low temperature, 
which is probably dominated by the conductivity of the Cu-O 
chains that is expected to be diminished at low $T$ (because of 
the strong tendency of the one-dimensional system to localization 
in the presence of disorder); on the other hand, at dopings lower 
than $y \simeq 6.55$, $\rho_a/\rho_b$ was found to {\it grow} 
with lowering temperature, which can be best attributed to the 
nematic charge stripes \cite{Kivelson} globally along the $b$ 
axis. In Fig. 3(b), we plot (using the data of Ref. 
\cite{anisotropy}) the difference in $\rho_a/\rho_b$ between 100 
K and 300 K, which indicates whether the anisotropy grows or 
diminishes upon lowering temperature; when $\rho_a/\rho_b(100 
{\rm K}) - \rho_a/\rho_b(300 {\rm K})$ is positive, that means 
that the anisotropy grows with lowering $T$ and thus is an 
indication that the stripes are becoming prominent in the charge 
transport. One can easily see that the boundary that separates 
the chain-dominated and stripe-dominated regimes is located 
around $y \simeq 6.55$ and this boundary corresponds exactly to 
the MIC determined from the $\kappa(H)$ behavior.

This comparison naturally suggests that the low-doping side of 
the MIC (``insulator" regime) is where the stripe correlations 
are pronounced. In this regime, it is possible that the magnetic 
field can enhance the static ordering of the stripes (as was 
suggested by neutron \cite{Lake,Khaykovich} and STM 
\cite{Hoffman} measurements) and causes the electrons to be 
localized. It is useful to note that the critical doping for QPT 
in YBCO, $p \simeq 0.07$, is considerably lower than that in 
other cuprates; it is at $p \simeq 1/8$ in BSLCO and at $p \simeq 
0.16$ in LSCO. This implies that, although the 
magnetic-field-induced static stripes are probably the common 
origin for the electron-localized state in the cuprates, the 
impact of the stripe correlations differs between the cuprates 
and such difference causes the variation in the location of the 
MIC. 

In summary, we show that the magnetic-field dependence of the 
thermal conductivity of YBa$_2$Cu$_3$O$_y$ single crystals 
measured at subkelvin temperatures indicates that the 
metal-to-insulator crossover under high magnetic field takes 
place at $y \simeq 6.55$. Since the previous measurements of the 
in-plane resistivity anisotropy indicated \cite{anisotropy} that 
the stripe correlations are pronounced in the transport below $y 
\simeq 6.55$, the present result gives strong support to the 
conjecture that existence of the stripe correlations 
characterizes the ``insulating" normal state of the cuprates.

\begin{acknowledgments} 
We thank A. N. Lavrov and J. Takeya for helpful discussions.  
\end{acknowledgments}

\end{document}